\documentclass[aps,prb,twocolumn,showpacs,showkeys,superscriptaddress]{revtex4}

\usepackage{amsmath}
\usepackage{graphics}

\include{MyMc}
\include{we}

\newcommand{\jAbsAv}{\ensuremath{j_\mathrm{AbsAv}}}
\renewcommand{\dj}{\ensuremath{\delta}}

\begin{document}

\title{Fractional vortex in asymmetric 0-$\pi$ long Josephson junctions}

\author{E. Goldobin}
\author{R. Kleiner}
\author{D. Koelle}
\affiliation{%
  Physikalisches Institut and Center for Collective Quantum Phenomena in LISA$^+$,
  Universit\"at T\"ubingen, Auf der Morgenstelle 14, D-72076 T\"ubingen, Germany
}

\date{%
  \today\ File: \textbf{\jobname.\TeX}
}

\begin{abstract}
  We consider an infinitely long 0-$\pi$ Josephson junction consisting of $0$ and $\pi$ regions having different critical current densities $j_{c,0}$ and $j_{c,\pi}$. The ground state of such a junction corresponds to a spontaneosly formed asymmetric semifluxon with tails decaying on different length scales. We calculate the depinning current of such a fractional vortex and show that it is different for positive and negative bias polarity. We also show that upon application of a bias current, the fractional flux (topological charge) associated with the vortex changes. We calculate the range of fractional flux associated with the vortex when the bias changes from negative to positive critical (depinning) values.
\end{abstract}



\maketitle

\section{Introduction}
\label{Sec:Intro}

A 0-$\pi$ Josephson junction (JJ) combines the properties of a conventional 0 JJ with critical current density $j_{c,0}>0$ with the properties of a $\pi$ JJ with $j_{c,\pi}<0$. Such JJs play an important role in the determination of the order parameter symmetry in novel superconductors\cite{Smilde:ZigzagPRL,Ariando:Zigzag:NCCO,Guerlich:2009:LTSEM-zigzag}, allow experiments with fractional Josephson vortex matter\cite{Hilgenkamp:zigzag:SF,Kirtley:2005:AFM-SF,DellaRocca:2005:0-pi-SFS:SF,Dewes:2008:ReArrangeE} and recently were used as a way to construct a $\varphi$ JJ with a tunable current-phase relation\cite{Sickinger:2012:varphiExp}.

Nowadays there are several technologies that allow to fabricate 0-$\pi$ long Josephson junctions (LJJs)\cite{Smilde:ZigzagPRL,Ariando:Zigzag:NCCO,Hilgenkamp:zigzag:SF,Scharinger:2012:ramp-zigzag:Ic(H),Weides:2006:SIFS-0-pi,Weides:2010:SIFS-jc1jc2:Ic(H),Kemmler:2010:SIFS-0-pi:Ic(H)-asymm}. One of them\cite{Weides:2006:SIFS-0-pi,Weides:2010:SIFS-jc1jc2:Ic(H),Kemmler:2010:SIFS-0-pi:Ic(H)-asymm} is based on employing a ferromagnetic barrier, which has a different thickness of the ferromagnet in the 0 and $\pi$ part and, correspondingly, different critical current densities $j_{c,0}$ and $j_{c,\pi}$. In practice, it is very difficult to control the thicknesses of the barrier in the 0 and $\pi$ region very precisely. Therefore, $j_{c,0}$ and $j_{c,\pi}$ are always different by absolute value, \ie, $j_{c,0}\neq|j_{c,\pi}|$. Moreover, for some devices such as $\varphi$ JJs\cite{Buzdin:2003:phi-LJJ} or, more general, for JJs with a tunable current-phase relation\cite{Goldobin:2011:0-pi:H-tunable-CPR,Sickinger:2012:varphiExp}, it is even necessary to make $j_{c,0}$ and $|j_{c,\pi}|$ different to achieve the required properties. However, most of the theoretical works so far deal with the idealized situation $j_{c,0}=|j_{c,\pi}|$.

Therefore, in this paper we consider an infinitely long 0-$\pi$ JJ with $j_{c,0}\neq|j_{c,\pi}|$ and investigate the qualitative differences in comparison with the symmetric case.

The paper is organized as follows. In Sec.~\ref{Sec:Asym-0-pi-LJJ.FracVort} we investigate the ground state and show that it corresponds to a semifluxon (fractional Josephson vortex carrying the flux $\pm\Phi_0/2$), which is pinned at the 0-$\pi$ boundary\cite{Xu:SF-shape,Goldobin:SF-Shape}, but has asymmetric tails. At non-zero bias current the shape of the fractional vortex can be obtained only numerically. Similar to the case of a symmetric 0-$\pi$ LJJ, by applying a large enough bias current, one can reach the depinning current\cite{Goldobin:SF-ReArrange,Malomed:2004:ALJJ:Ic(Iinj),Goldobin:2004:F-SF}, investigated in detail in Sec.~\ref{Sec:Ic}. We will see that in an asymmetric 0-$\pi$ JJ the critical current is different for positive and negative bias polarity. Moreover, we identify two different mechanisms of switching to the non-zero voltage state: depinning of the fractional vortex and primitive switching, \ie, when the bias current density exceeds $j_{c,0}$ or $|j_{c,\pi}|$. Further, in Sec.~\ref{Sec:Flux} we show that the fractional magnetic flux localized at the 0-$\pi$ boundary can deviate from $\pm\Phi_0/2$ when a bias current is applied. Finally, Sec.~\ref{Sec:Summary} summarizes this work.

\section{Fractional vortex in an asymmetric 0-$\pi$ LJJ}
\label{Sec:Asym-0-pi-LJJ.FracVort}

Consider an infinite asymmetric 0-$\pi$ LJJ with critical current density
$j_{c,0}>0$ in the 0 half ($x<0$) and $j_{c,\pi}<0$ in the $\pi$ half ($x>0$). For theoretical analysis it is convenient to work in normalized units. Therefore we introduce normalized critical current densities $\gamma_0=j_{c,0}/\jAbsAv$ and $\gamma_\pi=j_{c,\pi}/\jAbsAv$ with $\jAbsAv=(j_{c,0}+|j_{c,\pi}|)/2\geq 0$. Then the normalized critical current current along the 0-$\pi$ LJJ is given by
\begin{equation}
  j_c(x) =
  \begin{cases}
    \gamma_0>0 & x<0\\
    \gamma_\pi<0 & x>0
  \end{cases}
  . \label{Eq:j_c(x)}
\end{equation}
A normalized Ferrel-Prange equation describing the static solutions for the Josephson phase $\phi(x)$ reads
\begin{equation}
  \phi_{xx} - j_c(x)\sin\phi = -\gamma
  . \label{Eq:FP.Norm}
\end{equation}
Here $\gamma$ is the bias current density normalized to $\jAbsAv$, subscripts $xx$ denote the second derivative with respect to $x$, while the coordinate $x$ is normalized to
\begin{equation}
  \lambda_{J}(\jAbsAv) = \sqrt{\frac{\Phi_0}{2\pi\mu_0 d' \jAbsAv}}
  , \label{Eq:Def:lambda_J}
\end{equation}
calculated using the critical current density $\jAbsAv$. In Eq.~\eqref{Eq:Def:lambda_J} the quantity $\mu_0 d'$ is the inductance (per square) of the superconducting electrodes forming the LJJ.

For further analysis it sometimes will be convenient to introduce the deviation $\dj$ of $\gamma_0$ and $\gamma_\pi$ from the symmetric case ($\pm1$ in our normalized units), \ie,
\begin{eqnarray}
  \gamma_0 = 1+\dj, \quad \gamma_\pi = -1+\dj
  . \label{Eq:def:dj}
\end{eqnarray}
Thus one can use a single asymmetry parameter $|\dj|<1$ instead of $\gamma_0$ and $\gamma_\pi$.

For $\gamma=0$ the semifluxon solution of Eq.~\eqref{Eq:FP.Norm} is given by the fluxon tails properly jointed at the 0-$\pi$ boundary, \ie,
\begin{equation}
  \phi(x) =
  \begin{cases}
    4\arctan\left\{ \exp\left( \ratio{x-x_0  }{\lambda_{0  }} \right) \right\},      & x<0\\
    \\
    4\arctan\left\{ \exp\left( \ratio{x-x_\pi}{\lambda_{\pi}} \right) \right\} -\pi, & x>0
  \end{cases}
  , \label{Eq:mu(x):general}
\end{equation}
where $\lambda_0=1/\sqrt{\gamma_0}$ and $\lambda_\pi=1/\sqrt{|\gamma_\pi|}$ are normalized local Josephson penetration depths in the 0 and $\pi$ parts. They are given in units of $\lambda_J(\jAbsAv)$. The phase $\phi(0)$ and its derivative ($\propto$ magnetic field) $\phi_x(0)$ must be continuous at $x=0$, \ie,
\begin{eqnarray}
  \arctan(z_0) &=& \arctan(z_\pi)-\frac{\pi}{4}
  ; \label{Eq:BC0:phase}\\
  \frac{1}{\lambda_{0}}\frac{z_0}{1+z_0^2}&=&\frac{1}{\lambda_{\pi}}\frac{z_\pi}{1+z_\pi^2}
  , \label{Eq:BC0:field}
\end{eqnarray}
where
\begin{equation}
  z_0  = \exp\left( \frac{-x_0  }{\lambda_{0  }}     \right)
  ,\quad
  z_\pi= \exp\left( \frac{-x_\pi}{\lambda_{\pi}}     \right)
  .\label{Eq:Def:z}
\end{equation}
By taking $\tan(\ldots)$ of both sides of Eq.~\eqref{Eq:BC0:phase} we get
\begin{eqnarray}
  z_0 &=& \frac{z_\pi-1}{z_\pi+1}
  . \label{Eq:z0(z_pi)}
\end{eqnarray}
Solving Eqs.~\eqref{Eq:BC0:field} and \eqref{Eq:z0(z_pi)} for $z_0$ and $z_\pi$ we obtain two roots (for each of them). The negative roots can be neglected since $z_0$ and $z_\pi$ are positive by definition \eqref{Eq:Def:z}. Thus, the remaining positive roots are

\begin{subequations}
  \begin{eqnarray}
    z_\pi = \sqrt{\frac{|\gamma_\pi|}{\gamma_0}+1} + \sqrt{\frac{|\gamma_\pi|}{\gamma_0}} \geq 1
    ; \label{Eq:Sol:z_pi}\\
    z_0   = \sqrt{\frac{\gamma_0}{|\gamma_\pi|}+1} - \sqrt{\frac{\gamma_0}{|\gamma_\pi|}} \leq 1
    . \label{Eq:Sol:z_0}
  \end{eqnarray}
  \label{Eq:Sol:z}
\end{subequations}
From here, taking into account definitions \eqref{Eq:Def:z}, we get
\begin{subequations}
  \begin{eqnarray}
    x_\pi = -\frac{1}{\sqrt{|\gamma_\pi|}} \ln
    \left( \sqrt{\frac{|\gamma_\pi|}{\gamma_0}+1} + \sqrt{\frac{|\gamma_\pi|}{\gamma_0}} \right) \leq0
    ; \label{Eq:Sol:x_pi}\\
    x_0   = -\frac{1}{\sqrt{\gamma_0}} \ln
    \left( \sqrt{\frac{\gamma_0}{|\gamma_\pi|}+1} - \sqrt{\frac{\gamma_0}{|\gamma_\pi|}} \right) \geq0
    . \label{Eq:Sol:x_0}
  \end{eqnarray}
  \label{Eq:Sol:x}
\end{subequations}
%

%
\begin{figure*}[!tb]
  \includegraphics{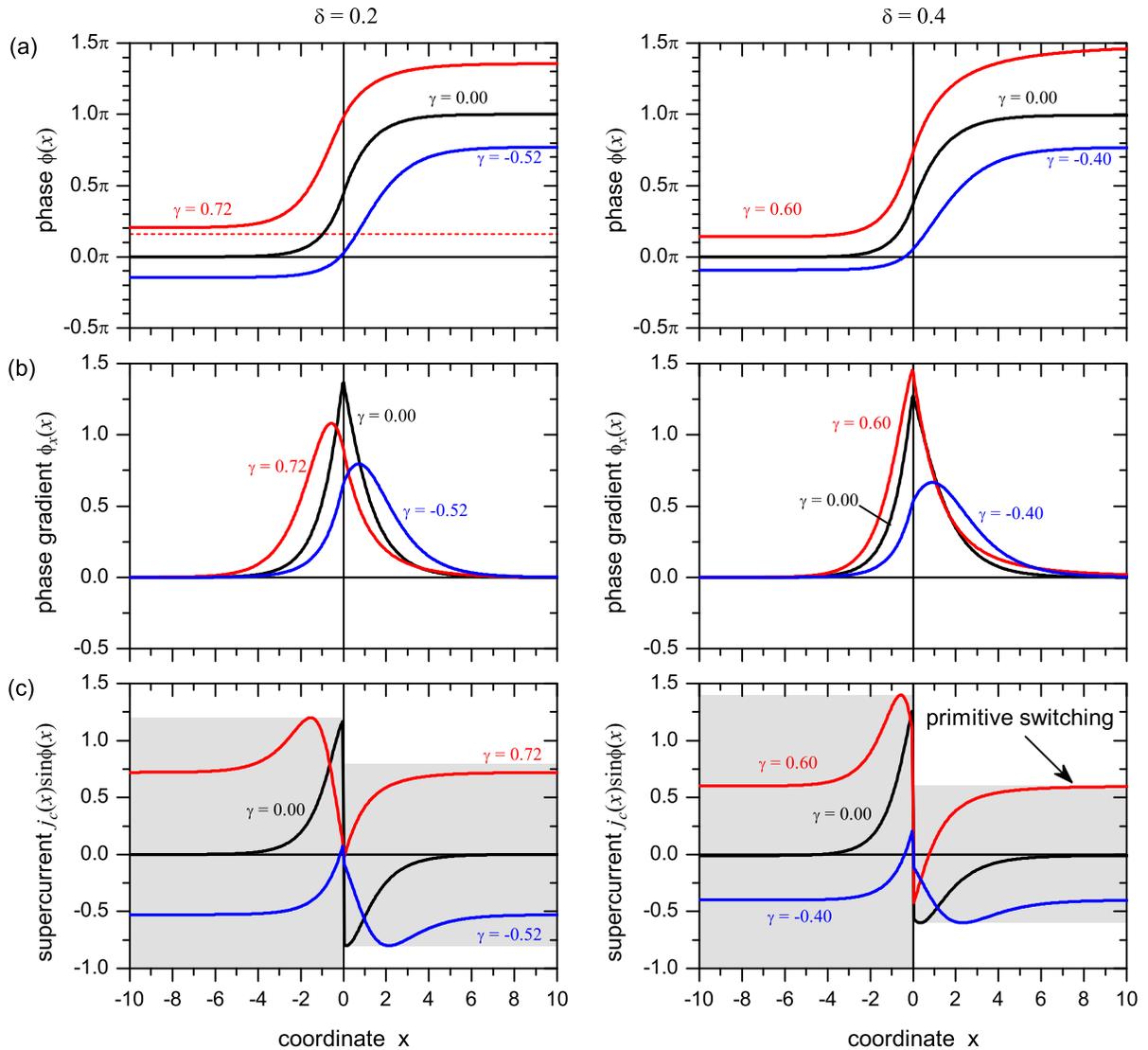}
  \caption{(Color online)
    The phase $\phi(x)$ (a), the magnetic field $\phi_x(x)$ (b) and the supercurrent $j_{0,\pi}\sin\phi(x)$ (c) of a postive vortex in a 0-$\pi$ LJJ with the critical current density asymmetry $\dj=0.2$ (left pannel), corresponding to depinning, and $\dj=0.4$ (right panel), corresponding to promitive switching. Profiles are shown for zero bias, positive precritical bias, and negative precritical bias. The gray areas in (c) indicate the critical current densities $\pm\gamma_0$ and $\pm\gamma_\pi$ in 0 and $\pi$ parts.
  }
  \label{Fig:Profiles}
\end{figure*}

For $\gamma \neq 0$ the static solutions of Eq.~\eqref{Eq:FP.Norm} cannot be obtained analytically. Therefore, Fig.~\ref{Fig:Profiles} shows profiles of the phase $\phi(x)$, magnetic field $\phi_x(x)$ and supercurrent $j_c(x)\sin\phi(x)$ obtained numerically. For zero bias current the profile coincides with the one given by Eq.~\eqref{Eq:mu(x):general} with $x_0$ and $x_\pi$ given by Eq.~\eqref{Eq:Sol:x}.

\section{Critical current in an asymmetric 0-$\pi$ LJJ}
\label{Sec:Ic}

\subsection{Depinning vs. primitive switching}
\label{Sec:Depin}

Althouh for $\gamma\neq 0$ the analytical solutions for the fractional vortex shape cannot be obtained, it is still possible to obtain the range of $\gamma$ where such a static solution exists.  By multiplying Eq.~\eqref{Eq:FP.Norm} by $2\phi_x$ and integrating, one finds (separately for 0 and $\pi$ region)
\begin{equation}
  \phi_x = \pm
  \begin{cases}
    \sqrt{2 \left[ C_0   - \gamma_0   \cos\phi - \gamma\phi \right]} & x<0\\
    \sqrt{2 \left[ C_\pi - \gamma_\pi \cos\phi - \gamma\phi \right]} & x>0
  \end{cases}
  , \label{Eq:phi'(phi)}
\end{equation}
where the integration constants $C_0$ and $C_\pi$ can be found from the boundary conditions at
\begin{eqnarray}
  \phi_x(\pm\infty) &=& 0
  ; \label{Eq:BC:phi'}\\
  \phi(-\infty) &=& \arcsin(\gamma/\gamma_0)
  ; \label{Eq:BC:phi@-infty}\\
  \phi(+\infty) &=& \pi-\arcsin(\gamma/\gamma_\pi)
  . \label{Eq:BC:phi@+infty}
\end{eqnarray}
Conditions \eqref{Eq:BC:phi@-infty} and \eqref{Eq:BC:phi@+infty} assume a vortex of positive polarity. For $C_0$ and $C_\pi$ one gets
\begin{eqnarray}
  C_0(\gamma)   &=& \gamma \arcsin\left( \frac{\gamma}{\gamma_0} \right)  + \sqrt{\gamma_0^2-\gamma^2}
  ; \label{Eq:C_0}\\
  C_\pi(\gamma) &=& \gamma \left[ \pi-\arcsin\left( \frac{\gamma}{\gamma_\pi} \right) \right]
  - \sqrt{\gamma_\pi^2-\gamma^2}
  . \label{Eq:C_pi}
\end{eqnarray}

\begin{figure}[!tb]
  \includegraphics{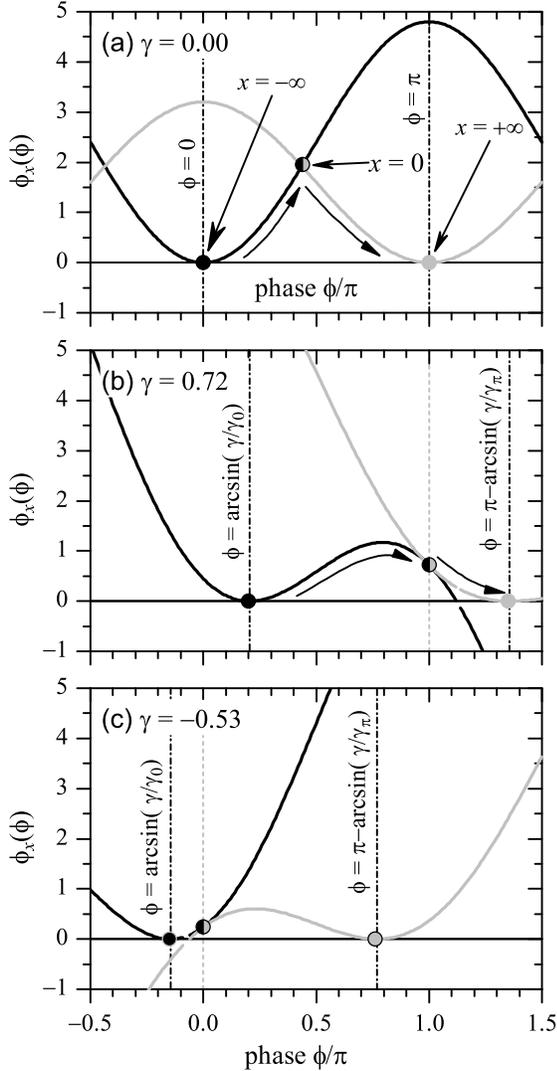}
  \caption{%
    The trajectories $\phi_x(\phi)$ on a phase plane for 0 segment (black) and for $\pi$ segment (gray) for asymmetry parameter $\dj=0.2$.
  }
  \label{Fig:Depin:PhasePlane}
\end{figure}

In Fig.~\ref{Fig:Depin:PhasePlane} we plot the phase plane curves $\phi_x(\phi)$ given by \eqref{Eq:phi'(phi)} separately for the 0 and the $\pi$ region and for different values of the bias current $\gamma$. At $\gamma=0$, see Fig.~\ref{Fig:Depin:PhasePlane}(a), the semifluxon solution corresponds to a line on the phase plane $\phi$-$\phi_x$ starting at $\phi=\arcsin(\gamma/\gamma_0)$, $\phi_x=0$ at $x=-\infty$ (black dot), and going towards the crossing point of black and gray curves. At the crossing (half-black-half-gray dot) at $x=0$ we switch to the gray $\pi$-region curve and follow it up to the point $\phi=\pi-\arcsin(\gamma/\gamma_\pi)$, $\phi_x=0$ (gray dot). Upon increasing $\gamma$ the crossing point of the two trajectories above transforms into a touching point at the critical bias current density $\gamma_{c+}>0$, see Fig.~\ref{Fig:Depin:PhasePlane}(b). After this the trajectories disconnect and the static solution is lost. Similar things happen for negative bias current density, see Fig.~\ref{Fig:Depin:PhasePlane}(c) shown for $\gamma_{c-}<0$. At the touching point, \ie, at $\phi=\phi(0)$ and at $\gamma_c$ (it can be either $\gamma_{c+}$ or $\gamma_{c-}$), the $\phi_x$ as well as $d(\phi_x)/d\phi$ of both trajectories are equal, \ie,
\begin{eqnarray}
  & & C_0  (\gamma_c) - \gamma_0   \cos\phi(0) - \gamma_c\phi(0)\nonumber\\
  &=& C_\pi(\gamma_c) - \gamma_\pi \cos\phi(0) - \gamma_c\phi(0)
  ; \label{Eq:fn}\\
  && \gamma_0\sin\phi(0) = \gamma_\pi \sin\phi(0)
  . \label{Eq:dr}
\end{eqnarray}
From Eq.~\eqref{Eq:dr} we conclude that either $\phi(0)=\pi$ (for $\gamma_c>0$) or $\phi(0)=0$ (for $\gamma_c<0$). Then Eq.~\eqref{Eq:fn} becomes
\begin{widetext}
\begin{eqnarray}
  &&  \gamma_c\left[
    \arcsin\left( \frac{\gamma_c}{\gamma_0} \right)+\arcsin\left( \frac{\gamma_c}{\gamma_\pi} \right)
  \right] 
  + \sqrt{\gamma_0^2-\gamma_c^2} - \sqrt{\gamma_\pi^2-\gamma_c^2}
  = \pi \gamma_c \mp(\gamma_0-\gamma_\pi)
  , \label{Eq:j_c}
\end{eqnarray}
\end{widetext}
where the upper sign corresponds to the positive $\gamma_c=\gamma_{c+}>0$ and the lower sign to a negative $\gamma_c=\gamma_{c-}<0$.

Note that if $\gamma_0=-\gamma_\pi$, so that the \lhs of Eq.~\eqref{Eq:j_c} vanishes, one obtains the well known depinning current of a semifluxon $\gamma_c=\pm 2/\pi$. For the asymmetric case $\gamma_0 \neq -\gamma_\pi$ Eq.~\eqref{Eq:j_c} provides an implicit dependence of $\gamma_c$ on $\gamma_0$ and $\gamma_\pi$. Note, however, that for given $\gamma_0$ and $\gamma_\pi$, $\gamma_c$ must be sought in the domain $|\gamma_c|<\min(\gamma_0,|\gamma_\pi|)$, otherwise some of the square roots and $\arcsin$ functions in Eq.~\eqref{Eq:j_c} are undefined in the real domain.


In terms of $\dj$ Eq.~\eqref{Eq:j_c} reads
\begin{widetext}
\begin{equation}
  \gamma_c\left[ \arcsin\left( \frac{\gamma_c}{1+\dj} \right)-\arcsin\left( \frac{\gamma_c}{1-\dj} \right) \right]+
  \sqrt{(1+\dj)^2-\gamma_c^2}-\sqrt{(1-\dj)^2-\gamma_c^2}
  = \pi \gamma_c - 2\sgn(\gamma_c)
  , \label{Eq:gamma_c(delta)}
\end{equation}
\end{widetext}
and $\gamma_c$ should be searched in the domain $|\gamma_c|<\min(1+\dj,1-\dj)$. One may or may not have solutions in this interval. And the solution of Eq.~\eqref{Eq:gamma_c(delta)} (dis)appears exactly when
\begin{equation}
  |\gamma_c(\dj)|=\min(1+\dj,1-\dj)
  . \label{Eq:gamma_c:primitive}
\end{equation}
Thus, we can find a ``critical'' value $\dj_c$ when solution(s) (dis)appear.

First, for $\gamma_c \geq 0$ and assuming $\dj>0$ we substite $\gamma_c=1-\dj$ from Eq.~\eqref{Eq:gamma_c:primitive} into Eq.~\eqref{Eq:gamma_c(delta)} and get the following equation for $\dj_c$
\begin{equation}
  \left[ \arcsin\left( \frac{1-\dj_c}{1+\dj_c} \right) -\frac32\pi \right](1-\dj_c)+2\sqrt\dj_c+2=0
  , \label{Eq:delta_c>0}
\end{equation}
which can be solved numerically to give $\dj_c\approx 0.2606$. For $\dj<0$, we substitute $\gamma_c=1+\dj$ from Eq.~\eqref{Eq:gamma_c:primitive} into Eq.~\eqref{Eq:gamma_c(delta)} and obtain an equation, which has only the trivial solution $\dj=-1$. Thus, for $\gamma_c \geq 0$ Eq.~\eqref{Eq:gamma_c(delta)} has a single solution only in the interval $\dj=-1\ldots\dj_c$.

Second, for $\gamma_c \leq 0$, similarly to the previous case we obtain that Eq.~\eqref{Eq:gamma_c(delta)} has a single solution, if $\dj=-\dj_c \ldots +1$.

Finally, when the asymmetry $\dj$ is such that Eq.~\eqref{Eq:gamma_c(delta)} delivers no solution for $\gamma_c$ (no vortex depinning current) for positive or negative bias polarity, the switching to the voltage state takes place at $\gamma_c$ given by Eq.~\eqref{Eq:gamma_c:primitive}, \ie, when the critical current density in the 0 or $\pi$ part will be exceeded. This can be written as
\begin{equation}
  \gamma_{c+} =
  \begin{cases}
    \mathrm{PositiveSolutionOf}\eqref{Eq:gamma_c(delta)} &\text{for } -1<\dj<\dj_c\\
    1-\dj &\text{for } \dj_c<\dj<+1
  \end{cases}
  , \label{Eq:gamma_c+.total}
\end{equation}
and, similarily,
\begin{equation}
  \gamma_{c-} =
  \begin{cases}
    -1-\dj &\text{for } -1<\dj<-\dj_c\\
    \mathrm{NegativeSolutionOf}\eqref{Eq:gamma_c(delta)} &\text{for } -\dj_c<\dj<+1
  \end{cases}
  . \label{Eq:gamma_c-.total}
\end{equation}
\begin{figure}[!htb]
  \includegraphics{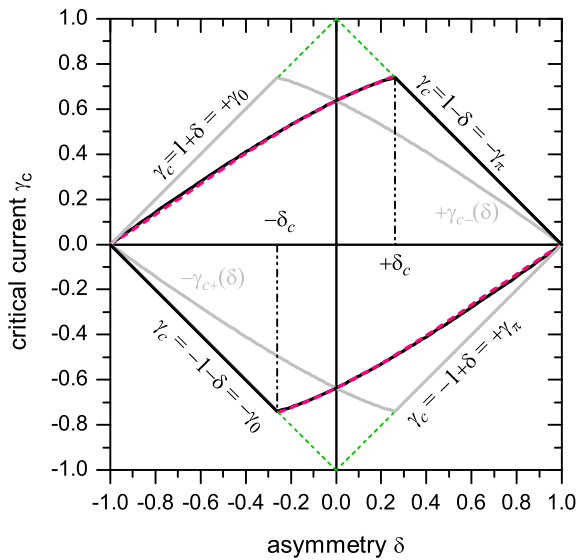}
  \caption{(Color online)
    The dependence of the normalized critical current $\gamma_{c\pm}$ of an asymmetric 0-$\pi$ LJJ on the asymmetry parameter $\dj$. The fat black continuous line shows the final $\gamma_{c\pm}(\dj)$ dependence for all $\dj$ given by Eqs.~\eqref{Eq:gamma_c+.total} and \eqref{Eq:gamma_c-.total}. It consists of vortex depinning branches (curved) given by Eq.~\eqref{Eq:gamma_c(delta)} and primitive switching branches (thin dash, straight part of fat black continuous line) given by Eq.~\eqref{Eq:gamma_c:primitive}. The approximation Eq.~\eqref{Eq:ApproxSol:gamma_c(delta)} is shown by the fat dashed line. Gray curves marked as $-\gamma_{c-}(\dj)$ and $-\gamma_{c+}(\dj)$ represent the critical current of the fractional antivortex.
  }
  \label{Fig:gamma_c(delta)}
\end{figure}

Fig.~\ref{Fig:gamma_c(delta)} shows the resulting dependence $\gamma_c(\dj)$ for positive and negative currents. One can see that in the general case $\dj\neq0$ the positive and negative critical (depinning) currents are not equal by absolute value. For $\dj=\pm\dj_c$ the pinning has a maxumim in one of the bias directions. For $|\dj|=1$, \ie, when one of the two halves has no Josephson properties (zero critical current density) the critical current vanishes.

It turns out that one can find a very good analytical approximation to the solution $\gamma_c(\dj)$ of the transcendental Eq.~\eqref{Eq:gamma_c(delta)}. If $\dj\ll1$, one can solve Eq.~\eqref{Eq:gamma_c(delta)} by Taylor expanding it near $\dj=0$ up to terms $\sim \dj$. One obtains
\begin{equation}
  \pi \gamma_c \mp 2 = 2\dj\sqrt{1-\gamma_c^2}
  . \label{Eq:gamma_c(delta).lin}
\end{equation}
The solution of this equation is
\begin{equation}
 \gamma_{c\pm} = 2\frac{\dj\sqrt{4\dj^2-4+\pi^2}\pm\pi}{4\dj^2+\pi^2}
  . \label{Eq:ApproxSol:gamma_c(delta)}
\end{equation}
The formula \eqref{Eq:ApproxSol:gamma_c(delta)} gives a very good approximation to the $\gamma_c(\dj)$ dependence obtained numerically by solving Eq.~\eqref{Eq:gamma_c(delta)}, as can be seen in Fig.~\ref{Fig:gamma_c(delta)}

Finally, we have checked our results by numerically solving a time-dependent sine-Gordon equation using \textsc{StkJJ}\cite{StkJJ}. A perfect agreement between the theory developped above and direct numerical simulation is found, including a crossover between different branches. Numerical results are not shown in Fig.~\ref{Fig:gamma_c(delta)} to avoid overcrowding. In Fig.~\ref{Fig:Profiles} we also show the profiles of the phase $\phi(x)$, phase gradient $\phi_x(x)$ ($\propto$ magnetic field) and supercurrent $j_c(x)\sin\phi(x)$ at precritical bias currents density. One can see that just before depinning ($\dj=0.2$ and $\gamma_c>0$) the second deriative of the phase (supercurrent) becomes continuous and has the same sign everywhere. On the other hand, just before primitive switching ($\dj=0.4$ and $\gamma_c>0$) the discontinuity of the supercurrent persists. These precritical profiles of the supercurrent are exactly the ones that should be seen in supercurrent distribution images obtained by low temperature scanning electron microscopy\cite{Guerlich:2009:LTSEM-zigzag,Guerlich:2010:SIFS-0-pi:LTSEM}.

The numerical simulations also reveal the branch corresponding to the critical current of a fractional antivortex. Indeed, the theoretical description presented above applies to a fractional vortex with positive flux (topological charge), see Eqs.~\eqref{Eq:BC:phi@-infty} and \eqref{Eq:BC:phi@+infty}. For an ``antivortex'' with negative flux (polarity) the situation reduces to the one considered above if one inverts the sign of the bias current. Thus, for an antivortex, the positive critical current is $-\gamma_{c-}(\dj)$, while the negative one is $-\gamma_{c+}(\dj)$. These curves are shown in Fig.~\ref{Fig:gamma_c(delta)} as well. One can see that, \eg, for $\dj<0$, the positive vortex becomes unstable at $\gamma_{c+}(\dj)$. It emits an integer fluxon and turns into a fractional ``antivortex'', which is still stable at this value of bias current. Depending on parameters such as damping, one will observe either the critical current of a fractional vortex $\gamma_{c+}(\dj)$ or the one of the ``antivortex'' $-\gamma_{c-}(\dj)$.

\subsection{Dynamics of depinning}

We have studied numerically the dynamics of the switching to the resistive state and compared the case of depinning at $\dj=0.2<\dj_c$ with the case of primitive switching at $\dj=0.4>\dj_c$.

For $\dj<\dj_c$ the depinning at $\gamma_{c+}$ starts from flipping the fractional vortex into an ``antivortex'' and emission of a fluxon. Further dynamics depends on the stability of the ``antivortex'' at given bias. First, if the ``antivortex'' is unstable, which is the case if its critical current $-\gamma_{c-}<\gamma_{c+}$, \ie, at $\dj>0$, then the flipping of the fractional (anti)vortex and the emission of fluxons and antifluxons continues. This type of depinning dynamics was discussed earlier for symmetric LJJs\cite{Goldobin:SF-ReArrange}. Second, if the ``antivortex'' is stable, which is the case for $\dj<0$, then further flipping does not take place and the state of the system is defined by the destiny of the emitted fluxon. For large damping the emitted fluxon moves away and is absorbed at the far end of the LJJ so that (after this transient) the LJJ remains in the static situation with an ``antivortex'' trapped and one finally measures in experiment/simulation the critical current $-\gamma_{c-}(\dj)$ of an antivortex. For moderate damping the emitted fluxon reflects from the edge of the LJJ as an antifluxon and starts moving back and forth (colliding with the fractional ``antivortex'' near $x=0$ on every pass) similarly to the dynamics at a zero-field step so that one detects a critical current $\gamma_{c+}(\dj)$ and a step on the IVC at larger bias current, presumably up to $-\gamma_{c-}(\dj)$.

For $\dj>\dj_c$, depinning at $\gamma_{c+}$ starts from switching the ``weaker'' ($|\gamma_\pi|<\gamma_0$) $\pi$-part of the LJJ into the resistive state, which may look like the penetration of an avalanche of fluxons from the edge of the LJJ. One switches to a finite voltage state and detects $\gamma_{c+}(\dj)$ as a critical current. Similar types of switching dynamics is observed at the negative critical current.

\section{Flux localized at a 0-$\pi$ boundary}
\label{Sec:Flux}

From Eq.~\eqref{Eq:BC:phi@-infty} and \eqref{Eq:BC:phi@+infty} it follows that the topological charge of the vortex is equal to $\pi$ (by absolute value) only if $\gamma=0$ or $\gamma_0=|\gamma_\pi|$. Otherwise it is given by
\begin{eqnarray}
  \wp(\gamma) &=& \pi-\arcsin(\gamma/\gamma_\pi)-\arcsin(\gamma/\gamma_0) \nonumber\\
      &=& \pi-\arcsin\left( \frac{\gamma}{-1+\dj} \right)-\arcsin\left( \frac{\gamma}{1+\dj} \right)
  . \label{Eq:TopoCharge(j)}
\end{eqnarray}
Naively, from Eq.~\eqref{Eq:TopoCharge(j)}, one concludes that in the case when $|\gamma_\pi| \ll \gamma_0$ ($\dj\to1$) one has $\wp\to 3\pi/2$ at $\gamma \to \gamma_\pi = \dj-1$. In the opposite case $|\gamma_\pi| \gg \gamma_0$ ($\dj\to-1$) one has $\wp\to \pi/2$ at $\gamma \to \gamma_0 = 1+\dj$. However, this is not completely true.

\begin{figure}[!htb]
  \includegraphics{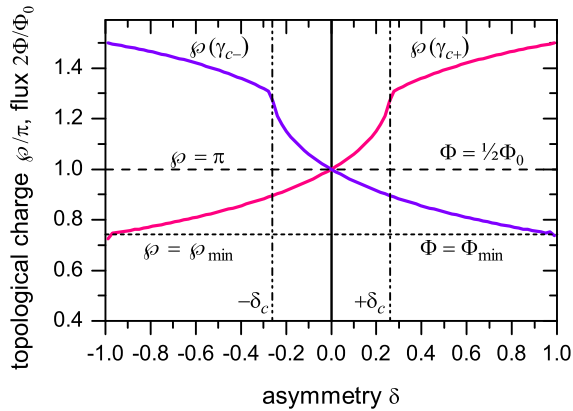}
  \caption{(Color online)
    The topological charge of the vortex $\wp$ at positive and negative critical current $\gamma_{c,\pm}$ as a function of asymmetry parameter $\dj$ calculated numetrically using Eq.~\eqref{Eq:TopoCharge(j)} with $\gamma=\gamma_c$ given by Eqs.~\eqref{Eq:gamma_c(delta)} and \eqref{Eq:gamma_c:primitive}.
  }
  \label{Fig:Flux}
\end{figure}

Since the topological charge given by Eq.~\eqref{Eq:TopoCharge(j)} is a monotonous function of $\gamma$ it reaches its extremum at $\gamma=\gamma_{c\pm}$. In Fig.~\ref{Fig:Flux} we plot the topological charge $\wp$ (the flux $\Phi=\Phi_0\wp/2\pi$) reached at the positive and negative critical current $\gamma_{c\pm}$ as a function of asymmetry $\dj$. When $\dj=0$ (symmetric case) the topological charge $\wp=\pi$ does not change with bias (see the crossing point of two curves). If $\dj\neq0$, the topological charge $\wp=\pi$ at $\gamma=0$ (horizontal dashed line) and changes in the interval between two curves upon application of bias. Each of the $\wp(\gamma_{c\pm})$ curves has a breaking point at $\pm\dj_c$, corresponding to a crossover from vortex depinning to primitive switching, see Fig.~\ref{Fig:gamma_c(delta)}. One can see that, indeed, the primitive switching branch gives the maximum topological charge $\wp_\mathrm{max}=3\pi/2$ ($\Phi_\mathrm{max}=3\Phi_0/4$). However, the depinning branch never reaches $\wp=\pi/2$. To calculate the minimum topological charge, one has to derive the limiting behavior of $\gamma_{c+}$ at $\dj\to-1$. By Taylor expanding Eq.~\eqref{Eq:gamma_c(delta)} near $\dj=-1$, we find that
\begin{equation}
  \gamma_{c+}\approx \alpha (\dj+1)\text{ at }\dj\to-1
  , \label{Eq:gamma_c+.behavior@delta-1}
\end{equation}
where $\alpha\approx0.7246$ is a solution of the trancendental equation
\begin{equation}
  1+\sqrt{1-\alpha^2}+(\arcsin\alpha-\pi)\alpha = 0
  . \label{Eq:f(alpha)=0}
\end{equation}
Substituting the expression \eqref{Eq:gamma_c+.behavior@delta-1} instead of $\gamma$ into Eq.~\eqref{Eq:TopoCharge(j)} for $\dj\to-1$ we obtain
\begin{equation}
  \wp_\mathrm{min} = \pi-\arcsin(\alpha) \approx 0.7418\pi
  . \label{Eq:wp_min}
\end{equation}
This corresponds to $\Phi_\mathrm{min}\approx 0.3709\Phi_0$ and is shown as the horizonal (short dashed) line in Fig.~\ref{Fig:Flux}.

At the end we would like to point out that the topological charge of an ``antivortex'' is, similar to Eq.~\eqref{Eq:TopoCharge(j)}, given by
\begin{eqnarray}
  \wp &=& -\pi-\arcsin(\gamma/\gamma_\pi)-\arcsin(\gamma/\gamma_0) \nonumber\\
      &=& -\pi-\arcsin\left( \frac{\gamma}{-1+\dj} \right)-\arcsin\left( \frac{\gamma}{1+\dj} \right)
  . \label{Eq:NSF:TopoCharge}
\end{eqnarray}
One can see that it is opposite to $\wp$ of the positive vortex only if $\gamma=0$ or $\gamma_0=|\gamma_\pi|$ ($\dj=0$). In the general case the topological charges (fluxes) of positive and negative fractional vortices at a given bias current are not equal. That is why we put the word ``antivortex'' in quotation marks throughout this paper.

\section{Summary}
\label{Sec:Summary}

We have demonstrated that an infinitely long 0-$\pi$ LJJ with unequal critical current densities $\gamma_0\neq |\gamma_\pi|$ in 0 and $\pi$ parts has a ground state corresponding to an asymmetric vortex of supercurrent carrying the magnetic flux $\pm\Phi_0/2$. The tails of this vortex decay on different length scales $\propto1/\sqrt{\gamma_0}$ and $\propto1/\sqrt{|\gamma_\pi|}$ as one goes away from the 0-$\pi$ boundary. Upon application of a small uniform bias current the vortex deforms but does not move away under the action of the Lorenz force as it is pinned at the 0-$\pi$ boundary. The fractional flux associated with the vortex in this state differs from $\pm\Phi_0/2$, as given by Eq.~\eqref{Eq:TopoCharge(j)}. This situation persists up to the critical bias current, which is different for positive and negative bias polarity, see Eqs.~\eqref{Eq:gamma_c+.total} and \eqref{Eq:gamma_c-.total}. Two mechanisms of switching to the resistive state are identified: (a) depinning of the vortex at the depinning current given by Eq.~\eqref{Eq:gamma_c(delta)} and (b) primitive switching when the bias current $\gamma$ exceeds $\gamma_0$ or $\gamma_\pi$ given by Eq.~\eqref{Eq:gamma_c:primitive}. The maximum deviation of the fractional flux from $\Phi_0/2$ is reached at the positive and negative critical currents. For the most asymmetric 0-$\pi$ junction (e.g. $\gamma_\pi=0$) the minimum possible flux $\Phi_\mathrm{min}\approx 0.3709\Phi_0$, while the maximum possible flux is $\Phi_\mathrm{max}=3\Phi_0/4$. Finally, the topological charges (magnetic fluxes localized near the 0-$\pi$ boundary) of positive and negative fractional vortices at a given bias current are not equal, except for the case of zero bias current.

\acknowledgments
We thank R.G. Mints and D. Heim for useful discussions. We acknowledge the financial support by the DFG (via projects GO-1106/5 and KO-1303/10).

\bibliography{SF,pi,software}


\end{document}